\documentstyle[11pt]{article}

\begin{document}
\newcommand{\BE}{\begin{equation}}
\newcommand{\EE}{\end{equation}}
\newcommand{\BA}{\begin{eqnarray}}
\newcommand{\EA}{\end{eqnarray}}
\newcommand{\LB}{\lambda_{\rm B}}
\newcommand{\PC}{\phi_{\rm c}}
\newcommand{\TG}{\tilde G}
\newcommand{\TE}{\tilde \eta}
\newcommand{\TF}{\tilde \phi}
\newcommand{\DE}{\bar\delta}
\begin{flushright}
{October 1992}
\end{flushright}
\vspace*{28mm}
\begin{center}
{\LARGE\bf Non-Gaussian Corrections to Higgs Mass\\[6mm]
           in Autonomous $\lambda \phi^4_{3+1}$}
\vspace{20mm}\\
{\Large Uwe Ritschel}
\vspace{13mm}\\
{\large\it
Fachbereich Physik, Universit\"at GH Essen, W-4300 Essen
(F R Germany)}
\vspace{27mm}\\
\end{center}
{\bf Abstract:} Recent calculations in one-loop and Gaussian
approximation,
using the so-called autonomous renormalization scheme, indicate a
comparatively massive, narrow Higgs excitation at about 2 TeV.
Here I show that this result qualitatively persists in the
framework of a post-Gaussian variational approximation for
the pure $O(N)$-symmetric $\phi^4$-theory for $N>1$.
The method is based on nonlinear transformations
of path-integral variables, and the optimization amounts
to a Schwinger-Dyson-type summation
of diagrams.
In the case of
$O(4)$, for example, I find $M_{\rm Higgs} = 2.3$
TeV, compared to
1.9 TeV and 2.1 TeV in 1-loop and Gaussian approximation,
respectively.
\vspace{4mm}
\newpage
\section{Introduction}
Five years ago Stevenson and Tarrach \cite{auto} suggested
certain
UV-flows for the bare parameters of the $\phi^4$-model which lead
to a finite Gaussian effective potential (GEP) \cite{stev0} in
$3+1$-dimensions.
The method was termed autonomous renormalization scheme (ARS) because
it is
not directly related to the
perturbative renormalization, the essential novelty being an infinite
rescaling of the field. The autonomously renormalized GEP exhibits all
generic features necessary for the Higgs mechanism, but it
eludes
triviality in that it is not equivalent to the
tree-level potential.

Recently essential progress has been made in understanding
the theoretical basis of the
ARS and in
deriving predictions
for physical quantities from it.
In a series of papers Consoli et al. \cite{cons} investigated
the ARS with renormalization group methods. They found asymptotic
freedom
in the asymmetric phase. Further, they discovered that the ARS also
works in the loop approximation on the one-loop level and that definit
predictions for the Higgs mass can be obtained under the natural
assumption that the theory has a massless particle in the symmetric
phase \cite{bran}.

Compared with standard methods, the ARS predicts a relatively
high Higgs mass at about 2 TeV, varying in the 10\%-range, depending
on whether the 1-loop potential or the GEP is considered, or whether
couplings of the scalars to gauge bosons are taken into account
or not \cite{rodr}. For instance for the $O(4)$-theory
in Gaussian approximation one finds
$M_{\rm Higgs}$= 2.05 TeV.

One of the most important questions which must be asked
in this context is whether
the autonomous ARS has a realization beyond
GEP and one-loop approximation. This is precisely the point which
is tackled in this paper with a post-Gaussian variational
calculation.
During recent years a method has
been developed, which allows to generalize the variational approach
in quantum field theory to non-Gaussian trial states in the
canonical formalism \cite{rits,iban} and non-Gaussian trial actions
in the path-integral formalism \cite{jens},
respectively. In the following I shall work in the covariant
setting
and restrict myself to the pure $O(N)$-theory.

\section{The method}
The generic (euclidean) action considered throughout the paper is
given by
\BE\label{acti}
S=\frac12 \int_x\left(-\partial_{\nu}\phi^i\partial_{\nu}\phi^i+
m_{\rm B}^2\phi^i\phi^i\right)
+ \LB\>\int_x\>(\phi^i\phi^i)^2\quad
\EE
with $\int_x=\int \>{\rm d}^4 x$.
In order to approximate the effective potential, $ V[\PC]$,
I introduce the nonlinear transformation
\BA \label{nonl}
\TF^1&=&\TE^1(p)+ \chi_0 \DE(p) +s\>
\int_{qr}\>c(q,r)\,\TE^a(q)\TE^a(r)\DE(p-q-r)\nonumber  \\
\TF^a&=&\TE^a(p) \qquad a=2,\ldots,N
\EA
which has Jacobian equal to unity,
where $\int_p = 1/(2\pi)^4\int\,{\rm d}^4p$,
$\>\DE(p)=(2\pi)^4\delta(p)$,
and tildes indicate Fourier amplitudes.

Transformation (\ref{nonl}) is appropriate for studying spontaneous
symmetry
breaking ${\rm O}(N)\rightarrow {\rm O}(N-1)$ and has previously been
used in the canonical approach \cite{iban}.
Leaving the path-integral measure invariant, (\ref{nonl}) is well
suited for a variational calculation, because it can be evaluated in
closed form, i.e.
does {\it not} lead to a series expansion of the expectation value.
The c-number $\chi_0$ and
the correlation function $s\>c(q,r)$ - the factor $s$ is split
off for normalization and as a bookkeeping device - are variational
parameters.

The upper bound on the effective potential derived from
(\ref{nonl}) is
\BE\label{aefa}
V(\PC) \le V_A(\PC) = \frac{1}{(\int_x)}\min\left.\left\{
-\log \,{\cal N}+ {\cal N}^{-1}
\int\>D\eta\>{\rm e}^{-S_G[\eta]}\>( S[\phi]-S_G[\eta] )\right\}
\right|_{\phi_{\rm c}}\quad ,
\EE
where the available variational parameters
have to be optimized
under the constraint
\BE\label{cons}
\phi_{\rm c}:={\cal N}^{-1}\int\>D\eta\>
{\rm e}^{-S_G[\eta]}\>\phi^1 = {\rm fixed},
\EE
{\it i.e.}, with the expectation value of the field
held fixed, and
\BE
{\cal N} = \int \>D\eta\>{\rm e}^{-S_G[\eta]}\>.
\EE
In the equations above the $\phi$'s have to be substituted with
transformation (\ref{nonl}).
$S_G$ is a quadratic test action,given by
\BE\label{tria}
S_G[\eta]=\frac12\>\int_{p}\left\{ \tilde\eta^1(p) \,G_L^{-1}(p)
\,\tilde\eta^1(-p)
+ \tilde\eta^a(p)\,G_T^{-1}(p)\,\tilde\eta^a(-p)\right\},
\EE
where $G_L$ and $G_T$ are adjustable propagators for
(with respect to the direction of symmetry breaking)
longitudinal and transversal modes, respectively.

\section{Autonomous Gaussian approximation}
If the nonlinear term is absent in (\ref{nonl}), the simple
identy $\PC=\chi_0$ holds and the optimization yields the (covariant)
GEP. (In contrast to the canonical formalism, this procedure can be
easily extended to the effective action \cite{jens}, but for my
present
purpose the effective potential is sufficient.)
In the following I am summarizing the basics of the ARS for
the Gaussian approximation,
and afterwards the analysis is extended to
the nonlinear transformations.

The bare
unoptimized GEP
is given by
\BA\label{gaus}
V_G &=& J^L +
     (N-1) J^T +
 \frac{1}{2} m_{\rm B}^2 \left[ I^L  + (N-1)I^T+ \phi_c^2 \right]
 \nonumber\\
& &+\LB \left[ 3(I^L+ \phi_c^2)^2 - 2 \phi_c^4 +
2(N-1)I^L(I^T + \phi_c^2) + (N^2-1)(I^T)^2 \right]\>,
\EA
with
\BE
J^{L(T)} = \frac12\int_p\>\log G_{L(T)}^{-1}
+\frac12 \int_p\>p^2 \,G_{L(T)}(p)
\EE
and
\BE
I^{L(T)}=\int_p\>G_{L(T)}(p)\>.
\EE

In the case of the GEP, the exact optimal
$G$'s in (\ref{tria}) have the form
\BE\label{gauf}
G_0(p;m)=\frac{ 1}{p^2+m^2}\>,
\EE
with optimization equations
\BE\label{caom}
\Omega^2 = m_{\rm B}^2 + 4 \LB \left( 3I_0(\Omega)+3\phi_c^2
+(N-1)I_0(\omega)
\right)
\EE
for longitudinal and
\BE\label{lcom}
\omega^2 = m_{\rm B}^2 + 4 \LB \left( I_0(\Omega)+\phi_c^2
+(N+1)I_0(\omega)
\right)
\EE
transversal mass \cite{onsy}.
With (\ref{gauf}) the $J$'s and $I$'s become
\BE
J(m) = -\frac12 \int_p \log G_0(p;m) - \frac12 m^2 I_0(m) \; , \qquad
I_0(m) = \int_p G_0(p;m)\; .
\EE

The ARS consists of the following UV-flows for the bare parameters:
\BE\label{auto}
m_{\rm B}^2 = \frac{m_0^2-12 \alpha I_0(0)}{I_{-1}(\mu)},\quad \LB =
\frac{\alpha}{I_{-1}(\mu)},\quad{\rm and}\quad \phi^2_c
= z_0 I_{-1}(\mu) \Phi^2\>,
\EE
where
\BE
I_{-1}(\mu) = 2 \int_p \left(G_0(p;\mu)\right)^2
\EE
is a
logarithmically divergent quantitiy which plays an inportant role in
the following,
and $\mu$ is an arbitrary renormalization mass, a dimensional
transmutation parameter for it replaces the dimensionless
coupling constant.
The second parameter, $m_0$, which turns out to be the mass at the
at the origin, is set to zero in the following \cite{rodr}.

Using (\ref{auto}) in (\ref{gaus}) one obtains
\BE\label{form}
V_G = \alpha I_{-1}(\mu)
\left( A \Omega^4 + B \omega^4 +
C\Omega^2\omega^2 + D\Omega^2 z_0\Phi^2 + E\omega^2
z_0\Phi^2 + z_0^2\Phi^4\right)
+ {\rm finite} + {\cal O}(1/I_{-1})
\EE
with
\BE\label{abcd}
A=\frac{1}{8\alpha} +\frac34, \quad B= \frac{N-1}{8\alpha} +
\frac{N^2-1}{4},
\quad C = 2 (N-1), \quad D=- 3, \quad E=-(N-1)\>,
\EE
i.e. all quartic and quadratic divergencies are cancelled, leaving
behind a logarithmically divergent expression.
In order to end up with a finite effective potential eventually,
also the logarithmically divergent term, proportional to $I_{-1}$,
has to vanish in the minimum with respect to the variational masses.
This amounts to an equation for $\alpha$ which can be solved
analytically.
The result is given by:
\BE
\alpha = \frac{1}{4 \,(1+\sqrt{N+3})}\>.
\EE
The remaining finite terms in (\ref{gaus}) are
\BE\label{fini}
V_G(\Phi) = \frac{1}{8\pi^2} z_0^2 \Phi^4 \left( X \log (\Phi^2/\mu^2)
-Y\right)
\EE
with
\BE\label{xval}
X = \frac{\sqrt{N+3}}{(2+\sqrt{N+3})(1+\sqrt{N+3})^2}\>.
\EE
For $X > 0$ the potential is stable, has massless excitations at
$\Phi=0$
and an asymmetric vacuum with massive particles.
(The constant $Y$ can be calculated but is not important for the
following
considerations.) The longitudinal mass - which is to be
with the Higgs mass
in the asymmetric minimum - is given by
\BE\label{mass}
\Omega^2 =  z_0 \sigma_L \Phi^2\quad {\rm with}
\quad \sigma_L=8\alpha\>.
\EE

It is worthwile to mention that the stability condition $X>0$ is
automatically satisfied, once the solution for $\alpha$ exists.
In (\ref{form}) all quantum corrections have to add up in order
to produce a negative contribution that cancels the positive
$z_0^2\Phi^4$.
The term proportional to $X$ is solely generated by
those subleading terms, which are
produced when the
longitudinal and transversal masses are eliminated in favor
of $\mu$.  The latter is done with the help of the relation
\BE\label{rela}
I_{-1}(m) = I_{11}(\mu) -
\frac{1}{8\pi^2} \log \frac{m^2}{\mu^2}\>,
\EE
and the
log-term yields the $\log (\Phi^2/\mu^2)$ in (\ref{fini}).
All other subleading terms contribute only to $Y$.
Thus, $X > 0$ is a direct consequence of the minus
sign in (\ref{rela}).

The finite factor $z_0$ that enters the wavefunction renormalization
in (\ref{auto}) can be calculated from the renormalization condition
\BE\label{znot}
\left.\frac{{\rm d}^2 V}{{\rm d} \Phi^2}\right|_{\Phi=\Phi_v}= \left.
\Omega^2\right|_{\Phi=\Phi_v}= z_0 \sigma_L \Phi_v^2\>,
\EE
where $\Phi_v$ is the value of the field in the asymmetric minimum.
This leads to
\BE
z_0 = \pi^2 \>\frac{\sigma_L}{X} = \pi^2\> \frac{10 + 6 \sqrt{N+3} +
2N}{\sqrt{N+3}}\>.
\EE
If now the well-known
value $\Phi_v = 0.246$  TeV is inserted in (\ref{mass}),
$M_{\rm Higgs}$ can be computed; the table below shows the results.

\section{Non-Gaussian contributions}
I am switching on the nonlinear term now, leaving the correlation
function $c(p,q)$ subject to optimization.
The expectation value can be calculated straightforwardly.
The unoptimized effective potential becomes
\BA
V_{A} &=& V_G+ \Delta V = V_G + s^2 \chi_7 + \frac{1}{2}
m_{\rm B}^2 s^2
\chi_2
+ \LB \left\{
4 \phi_c s \chi_3 + 2 s^2 \chi_5  \right. \\
& & \left.
+ s^2\left[ 6 I^L + 2 (N-1) I^T + 6 \phi_c^2 \right] \chi_2
 + 4 \phi_c s^3 \chi_4 + 3 s^4 (\chi_6 + \chi_2^2) \right
 \}
\EA
where $V_G$ is the Gaussian result, with yet undetermined
longitudinal and transversal propagators,
and
\BA
\chi_2 &=& 2 (N-1) \int_{pq} c^2(p,q) G_T(p) G_T(q),
\\
\label{chi3}
\chi_3 &=& 2 (N-1) \int_{pq} c(p,q) G_T(p)G_T(q),
\\
\chi_4 &=& 8 (N-1) \int_{pqr} c(p,q)c(p,r)c(q,-r)
 G_T(p)G_T(q)G_T(r),
\\
\chi_5 &=& 8 (N-1) \int_{pqr} c(p,q)c(q,r)G_T(p)G_T(q)G_T(r),
\\
\chi_6 &=& 16 (N-1) \int_{pqrs} c(p,q)c(p,r)c(q,s)c(r,s)
          G_T(p)G_T(q)G_T(r)G_T(s),
\\
\chi_7 &=& (N-1) \int_{pq} (p+q)^2 c^2(p,q) G_T(p)G_T(q).
\EA

It is the term linear in $s$, proportional to $\chi_3$, that assures
an improved energy compared to the GEP, at least
on the level of the bare theory.
Neglecting for the present the higher-order terms proportional to
$\chi_4$ and $\chi_6$ - the justification
for that will be given below -, the optimization can be carried out,
reducing
the problem to a Schwinger-Dyson-type integral equation for
the optimal correlation function $c(p,q)$ :
\BE\label{swdy}
c(p,q)
= G_0(p+q;\Omega)\left(1 -
8 \LB \int_r \left[ c(p,r) + c(q,r) \right]
G_T(r) \right),
\EE
with
\BE
\Omega^2 = m_{\rm B}^2 + 4 \LB \left(
 (N-1)I^T + 3 (I^L +s^2\chi_2+\phi_c^2 ) \right)
\EE
and where
\BE\label{sexa}
s= -4\LB \phi_c
\EE
has been chosen for convenient normalization.

The solution of (\ref{swdy}) leads to a net energy-decreasing term
\BE\label{nete}
\Delta V = - 8\LB^2\phi_c^2 \bar\chi_3+{\cal O}(s^4)
\EE
where $\bar\chi_3$ is (\ref{chi3}) with optimal correlator.
The $\chi$-integrals satisfy
\BE\label{sum}
2\bar\chi_7 +\Omega^2 \bar\chi_2 +4\LB\bar\chi_5
= \bar\chi_3 \>.
\EE

The iteration of (\ref{swdy}) reveals that (\ref{nete}) contains
an infinite
series of contributions to the two-point vertex function -
the first four graphs are depicted in Fig. 1 -,
thus demonstrating that the approximation goes
far beyond the Gaussian approximation
which sums up the ``cactus" graphs.

\section{Renormalization}
Dimensional analysis shows that the bare correction $\Delta V$ is
quadratically divergent. As a consequence, it
necessitates a correction to the
Gaussian mass counterterm in (\ref{auto}). As discussed in
\cite{reno},
however, any such correction to $m_{\rm B}^2$ multiplies
the whole mass term,
\BE
 \langle \phi^i\phi^i\rangle =
 \left(I^L+s^2\chi_2+\phi_c^2+(N-1)I^T\right)\>,
\EE
thus
generating new
(quartically and quadratically) divergent terms which have no direct
counterpart
in the expectation value. This dilemma originally led to the
notion of the instability of
variational calculations \cite{wudk} in general
and the ARS in particular \cite{inst}.
With the help of the diagrammatic representation, however, a deeper
understanding of the problem and, eventually, a
straightforward solution can be obtained.

The first diagram in the series Fig.1, the so-called barred circle,
has the close relative shown
in Fig. 2, a vacuum diagram that may be generated from the barred
circle by
attaching two external lines and connecting them. In second-order
$\delta$-expansion \cite{stan}, for example,
both diagrams are present and the renormalization
works as Fig. 2  contains the barred circle as a divergent
subgraph.
In the variational calculation, however,
the situation is quite different. Fig. 2
is not generated automatically. This can be verified
by considering the effective potential at the origin
where the optimization can be carried out exactly \cite{iban}.
Consequently, the counterterm for the renormalization of the barred
circle
causes (quartically) divergent subtractions which are actually
supposed to
renormalize vacuum graphs.

There are two possible solutions to this problem. Firstly, one can
think
of an ansatz which generates both diagrams. This can
in principle be achieved
by a transformation like (\ref{nonl})
with cubic nonlinear term of the form
$\eta^1 \eta^a\eta^a$. However, this transformation
has nontrivial Jacobian and thus makes the
evaluation of the ansatz much more complicated and the
applicability of
the variational method questionable. Secondly, as suggested in
\cite{reno},
one can choose a sub-optimal form of the variational parameters,
just generating the divergence, which otherwise would be caused by
Fig. 2 and which then can be subtracted by the mass counterterm.

The ansatz is given by
\BE\label{fans}
G_L(p)=G_0(p;\Omega)-s^2\xi(p)
\EE
with
\BE\label{xi}
\xi(p)\>=\>2 (N-1)\, G_T(p)
\>\int_q\>c^2(p,q)     G_T(q)
\EE
for the longitudinal propagator and
\BE
G_T(p)=G_0(p;\omega)
\EE
for the transversal propagator.

With (\ref{fans}) one obtains
\BE\label{iver}
I^L=I_0(\Omega)-s^2\chi_2
\EE
and
\BE\label{jver}
J^L=J(\Omega)+\frac12\Omega^2 s^2\chi_2+{\cal O}(s^4)
\EE

The next step is to set
\BE\label{anss}
s\>=\>-4 \LB\left(\phi_c-
\sqrt{-\Delta(\Omega)-(N-1)\Delta(\omega) }\right)\>
\EE
where
\BE\label{D}
\Delta(m)=I_0(m)-I_0(0) = -\frac12 m^2 I_{-1}(\mu) + \frac{1}{16\pi^2}
\left( \log \frac{m^2}{\mu^2} -1 \right) \,.
\EE
As opposed to (\ref{sexa}) the ansatz
(\ref{anss})
generates a term
proportional to
\BE
\left( \Delta(\Omega) +(N-1)\Delta(\omega)\right) \>\chi_3\>,
\EE
whereas on account of the
identity (\ref{sum}) terms containing the
square root of the $\Delta$'s cancel among each other.

Finally, by demanding that the excitation at
the origin remains massless, the
bare mass
\BE\label{mb}
m_{\rm B}^2=
-12 \LB \,I_0(0)
+ 16 \LB^2\left.\bar{\chi}_3\right|_{\phi_c=0}
\EE
is obtained.
As a result of (\ref{fans}), (\ref{anss}), and (\ref{mb})
the effective potential is given by
\BE\label{efba}
V_{A} = V_G - 8 \LB^2 \left(\bar\chi_3-
\left.\bar\chi_3\right|_0\right)\left(\Delta(\Omega)+
(N-1)\Delta(\omega) +\phi_c^2 \right)+ {\cal O}(\phi_c^4)\>.
\EE
This expression is not yet finite, but all power divergencies
have been removed.
As I shall demonstrate below, the only remaining divergencies are
logarithmic and, like in the GEP-case,
they can be cancelled by adjusting the parameter
$\alpha$ in $\LB$.

Before I proceed, the terms
${\cal O}(\phi_c^4)$ have to be discussed.
{}From (\ref{auto}) and (\ref{anss}) it
follows that $s^2 \propto 1/I_{-1}$. Analyzing the integrals which
occur
in ${\cal O}(\phi_c^4)$ (like $\chi_4$, $\chi_6$, and other
contributions
produced by (\ref{fans})), one finds that all these terms are at most
${\cal O}(1/I_{-1})$ and, thus, they are not affecting the
finite parts of the effective potential in the ARS.

\section{Evaluation of the effective potential}
In this section the Schwinger-Dyson
equation (\ref{swdy}) will be discussed.
The equation is in a sense the higher-order analogue to the
self-consistency equation of the Gaussian approximation.
The integral equation corresponding to the
latter is extremely simple and can be solved algebraically. With
(\ref{swdy}) the situation is more complicated.
The technical problems are closely related to the ones
occuring in the  $1/N$ expansion \cite{ma}.
Like there, subsets of of diagrams,
containing chains of bubbles of different
length, can in principle be summed up.
In the following I am not
proceeding along
those lines, however, because there are good reasons to
believe
that the series converges rapidely when the factor $\alpha$ (which
is the expansion parameter in the integral equation) assumes
its physical value, where
logarithmic divergencies are cancelled.

The leading (logarithmic) divergence of the diagrams generated
by (\ref{swdy}) is proportional to
$I_{-1}^k$ ,
where $k$ is the number of vertices. With the
help of (\ref{D}) one finds that
$\Delta V \propto I_{-1}$,
like in the Gaussian approximation.
I have calculated explicitely the contributions from the first
and second diagram in Fig. 1.
The leading divergence of these graphs is
given by
\BE\label{ledi}
-\frac18\>(\Omega^2 + 2 \omega^2) I_{-1}^{2} \quad{\rm and}\quad
-\frac23\>(\Omega^2 + 2 \omega^2) I_{-1}^{3}   \>,
\EE
respectively.

With the ansatz
\BE
\Omega^2 = \sigma_L z_0 \Phi^2\>,\quad\omega^2 =\sigma_T z_0 \Phi^2
\EE
the divergent part
of the effective potential again has the form (\ref{form}).
The coefficients read
\BA\label{coef}
A' &= &A- \Delta, \quad
B' = B - 2 (N-1) \Delta,\quad\nonumber\\
C' &=& C - (N+1)\Delta,\quad
D' = D + 4  \Delta\,\quad
E' = E + 2 \Delta \>
\EA
with
\BE
\Delta = (N-1) \alpha \left(1 - \frac{16}{3}\alpha + {\cal
O}(\alpha^2)\right)
\EE
and where the unprimed letters stand for the Gaussian coefficients
(\ref{abcd}).
One derives three equations
from the conditions that
the term proportional $I_{-1}$ vanishes in the
minimum with respect to the variational masses.
These equations can be solved numerically
giving $\alpha$, $\sigma_L$, and $\sigma_T$ as
functions of $N$.

Eventually also the finite parts of the
effective potential can be obtained. They again
have the form (\ref{fini}), and
the factor $X$ can be calculated on account of (\ref{rela}) directly
from the leading divergence.
The results for the Higgs mass are given
in the table.

\section{Discussion of results}
I have demonstrated that
within the $O(N)$- theory the autonomous renormalization scheme
exists
beyond  the one-loop and Gaussian variational
approximation. This result gives some confidence in the ARS and
makes it unlikely that one is dealing with an artifact of the
GEP.
The optimized nonlinear transformations
lead to a Schwinger-Dyson-type integral equation,
formally summing up a series of
diagrams that goes far beyond the one that is taken into account
by the Gaussian method.
The integral equation has been approximated
by the second iteration, which contains diagrams
with up to three loops.
An estimate of the
changes due to four-loops indicates that the
error of the approximation is less than 3\%.

The results for the Higgs mass
are shown in the table. For low values of $N$, the mass is higher
than in one-loop and Gaussian approximation.
For instance, for $N=4$ it lies
10\% above the Gaussian and
20\% above the one-loop prediction.
For higher $N$ it drops substantially below the Gaussian results.
\begin{center}
\begin{tabular}[r]{||p{1.2cm}|p{2.5cm}|p{2.5cm}|| }
\hline
N & Gaussian & Non-Gaussian\\ \hline
1 & 2.19 & $-$\\
2 & 2.13 & 2.21 \\
4 & 2.05 & 2.27 \\
7 & 1.97 & 2.22 \\
10 & 1.95 & 2.08 \\
20 & 1.84 & 1.70 \\
100 & 1.69 & 1.34\\ \hline
\end{tabular}
\end{center}
{\bf Table 1} {Values of Higgs mass (in TeV) for different values
of
$N$ from Gaussian and non-Gaussian approximation. } \\

The present calculation shows that the
error in the Higgs mass
due to the approximative treatment of the self interaction
lies in the same order of magnitude as
corrections due to interactions with gauge bosons and fermions.
Hence,
if the phenomenological consequences of the ARS are to be
taken seriously, one clearly has to confirm them by alternative
improved approximations, like the conventional loop expansion,
the $\delta$-expansion \cite{stan}, or the optimized expansion
\cite{okop}.

\newpage\noindent
{\Large\bf Figure captions}\vspace{1cm}\\
{\bf Fig. 1} Four contributions
in the series of diagrams generated by the Schwinger-Dyson
equation
(\ref{swdy}). Thin (thick) lines represent transversal (longitudinal)
propagators. \vspace{5mm}\\
{\bf Fig. 2} Vacuum diagram that contains the ``barred circle" as
divergent
subgraph.
\end{document}